\documentclass[twocolumn,showpacs,amsmath,amssymb]{revtex4}
\usepackage{graphicx}
\usepackage{dcolumn}
\usepackage{bm}
\newcommand{\be}{\begin{eqnarray}}
\newcommand{\en}{\end{eqnarray}}
\newcommand{\ben}{\begin{eqnarray*}}
\newcommand{\enn}{\end{eqnarray*}}
\newcommand{\pa}{\partial}

\newcommand{\f}{\frac}

\newcommand{\p}{\paragraph{}}
\newcommand{\bi}{\begin{itemize}}
\newcommand{\ei}{\end{itemize}}

\newcommand{\la}{\langle}
\newcommand{\ra}{\rangle}

\newcommand{\R}{\Rightarrow}
\renewcommand{\r}{\rho}
\renewcommand{\p}{\bot}
\renewcommand{\a}{\alpha}
\renewcommand{\b}{\beta}

\begin{document}
\title{On the use of Kolmogorov-Landau approach in deriving various correlation functions in 2-D incompressible turbulence}
\author{Sagar Chakraborty}
\email{sagar@bose.res.in}
\affiliation{S.N. Bose National Centre for Basic Sciences\\Saltlake, Kolkata 700098, India}
\date{November 20, 2006}
\begin{abstract}
We look at various correlation functions, which include those that involve both the velocity and the vorticity fields, in 2-D isotropic homogeneous decaying turbulence.
We adopt the more intuitive approach due to Kolmogorov (and subsequently, Landau in his text on fluid dynamics) and show that how the 2-D turbulence results, obtainable using other methods, may be established in a simpler way.
Also, some experimentally verifiable correlation functions in the dissipation range have been derived for the same system.
The paper also showcases the inability of the Kolmogorov-Landau approach to get the ``one-eighth law'' in the enstrophy cascade region.
As discussed in the paper, this may raise the spectre of logarithmic corrections once again in 2-D turbulence.
\end{abstract}
\maketitle
\section{INTRODUCTION}
Kolmogorov's four-fifths law\cite{Kolmogorov} is a landmark in the theory of turbulence because it is an exact result.
In three spatial dimensions, this law says that the third order velocity correlation function behaves as:
\ben
\left<\left[\left\{\vec{v}(\vec{x}+\vec{r})-\vec{v}(\vec{r})\right\}.\f{\vec{r}}{|\vec{r}|}\right]^3\right>=-\f{4}{5}\varepsilon r
\enn
where $\varepsilon$ is the rate per unit mass at which energy is being transferred through the inertial range.
The inertial range is the intermediate spatial region postulated by Kolmogorov where the large scale disturbances (flow maintaining mechanisms) and the molecular scale viscous dissipation play no part.
This result is of such central significance that attempts are regularly made to understand it afresh and to extend it in other situations involving turbulence.
There appears to be following two primary methods of obtaining this result:
\begin{enumerate}
\item
The original Kolmogorov method put forward in details in the fluid dynamics text due to Landau and Lifshitz\cite{Landau}.
There is no external forcing in this approach and the equality of dissipation rate and forcing rate for the energy is never enforced.
\item
A ``field-theoretic technique'' because of its analogy to anomalies in the calculations of certain field-theoretic correlation functions.
This is the path following in the more recent turbulence text by Frisch\cite{Frisch}.
In this approach, there is an external forcing that maintains a steady state turbulence.
\end{enumerate}
The two approaches yield the same result as they should.\\
If we consider the two-dimensional turbulence, then in the inviscid limit, we have two conserved quantities -- energy and enstrophy.
This gives rise to two fluxes with the enstrophy flux occurring from the larger to the smaller spatial scales.
The energy flux goes in the reverse direction.
Recently, Bernard\cite{Bernard} has used the second of the above mentioned techniques to obtain the third order structure function for both the energy and the enstrophy cascade regions.
Interestingly enough, he finds an exact answer for the two-point third order correlation function in the enstrophy cascade.
Now, the correlation functions in the enstrophy cascade are often plagued by logarithmic corrections\cite{Kraichnan} and in view of this Bernard's result is particularly interesting.
We believe that the issue is important enough that a derivation of Bernard's result using the Kolmogorov-Landau approach should be useful.
To the best of our knowledge and literature survey, such an approach remains unreported. 
So this is what we have attempted here and while the energy cascade result does come out in agreement with Bernard's, it appears that the Landau treatment cannot yield the prefactor of the so-called ``$1/8\textrm{th}$-law'' for the two-point third order correlation function in the enstrophy cascade dominated regime.
This would be the case if logarithmic corrections exist.
We believe that this raises the spectre of logarithmic corrections once again.
Besides, we also have derived some other correlation functions which deal with vorticity fields in the inertial region and also some two-point second order correlation functions in the dissipative region following the arguments of Landau, thereby consolidating the equivalence between the two approaches mentioned in the beginning. 
\section{SECOND ORDER CORRELATION FUNCTION FOR ENSTROPHY CASCADE IN DISSIPATION RANGE}
It is a well-established fact that there exists a direct-cascade of enstrophy in two-dimensional (2-D) turbulence.
One defines enstrophy as $\Gamma=\f{1}{2}\int_{\textrm{all space}}\omega^2 d^2\vec{\r}$ where $\omega=\pa_xv_y-\pa_yv_x$ is the vorticity in the Cartesian coordinates; $v$ being the velocity field.
As we shall consider incompressible fluids only ($\vec{\nabla}.\vec{v}=0$), we shall take density to be unity and let $\vec{\r}$ take over the task of representing position vector in 2-D plane.
The enstrophy flows through the inertial range and gets dissipated near dissipation scale.
Using the antisymmetric symbol $\varepsilon_{\a\b}$ that has four components, viz. $\varepsilon_{11}=\varepsilon_{22}=0$ and $\varepsilon_{12}=-\varepsilon_{21}=1$, one may define the mean rate of dissipation of enstrophy per unit mass as:
\be
&{}&\eta\equiv\nu\la \vec{\nabla}\omega.\vec{\nabla}\omega\ra\nonumber\\
\Rightarrow&{}&\eta=\nu\varepsilon_{\tau\a}\varepsilon_{\theta\b}\la (\pa_\tau\pa_\gamma v_{\a})(\pa_\theta\pa_\gamma v_{\b})\ra
\label{0}
\en
Here, angular brackets denote an averaging procedure which averages over all possible positions of points $1$ and $2$ at a given instant of time and a given separation.
Now, if $\vec{v}_1$ and $\vec{v}_2$ represent the fluid velocities at the two neighbouring points at $\r_1$ and $\r_2$ respectively, one may define rank two correlation tensor:
\be
B_{\a\b}\equiv\la(v_{2\a}-v_{1\a})(v_{2\b}-v_{1\b})\ra
\label{1}
\en
For simplicity, we shall take a rather idealised situation of turbulence flow which is homogeneous and isotropic on every scale, a case achievable in practice in a vigorously-shaken-fluid left to itself.
The component of the correlation tensor will obviously, then, be dependent on time, a fact which won't be shown explicitly in what follows.
As the features of local turbulence is independent of averaged flow, the result derived below is applicable also to the local turbulence at distance $\r$ much smaller than the fundamental scale.
Isotropy and homogeneity suggests following general form for $B_{\a\b}$\cite{Robertson}
\be
B_{\a\b}=A_1(\r)\delta_{\a\b}+A_2(\r)\r^o_\a\r^o_\b
\label{2}
\en
where $A_1$ and $A_2$ are functions of time and $\r$.
The Greek subscripts can take two values $\r$ and $\p$ which respectively mean the component along the radial vector $\r$ and the component in the transverse direction.
Einstein's summation convention will be used extensively.
Also,
\ben
\vec{\r}=\vec{\r}_2-\vec{\r}_1,\phantom{xxx}\r^o_\a\equiv\r_\a/{|\vec{\r}|},\phantom{xxx}\r^o_\r=1,\phantom{xxx}\r^o_\p=0
\enn
using which in the relation (\ref{2}), one gets:
\be
B_{\a\b}=B_{\p\p}(\delta_{\a\b}-\r^o_\a\r^o_\b)+B_{\r\r}\r^o_\a\r^o_\b
\label{3}
\en
One may break the relation (\ref{1}) as
\be
B_{\a\b}=\la v_{1\a}v_{1\b}\ra+\la v_{2\a}v_{2\b}\ra- \la v_{1\a}v_{2\b}\ra-\la v_{2\a}v_{1\b}\ra
\label{4}
\en
and defining
\be
b_{\a\b}\equiv\la v_{1\a}v_{2\b}\ra
\label{5}
\en
one may proceed, keeping in mind the isotropy and the homogeneity, to write
\be
B_{\a\b}=\la v^2\ra\delta_{\a\b}-2b_{\a\b}
\label{6}
\en
Again, having assumed incompressibility, one may write:
\be
&{}&\pa_\b B_{\a\b}=0\nonumber\\
\Rightarrow&{}& B'_{\r\r}+\f{1}{\r}(B_{\r\r}-B_{\p\p})=0\nonumber\\
\Rightarrow&{}&B_{\p\p}=\r B'_{\r\r}+B_{\r\r}
\label{7}
\en
where the equation (\ref{3}) has been used and prime ($'$) denotes derivative w.r.t. $\r$.
Near the dissipation region the flow is regular and its velocity varies smoothly which allows to expand $v$ in a series of power of $\r$.
One must take $v\sim \r^2$ neglecting the higher powers ($v\sim \r$ is not taken because it leads to the contradictory result that $\eta=0$ as can be seen from the relation (\ref{0})).
So, treating $a$ as a proportionality constant, let $B_{\r\r}=a\r^4$, which means $B_{\p\p}=5a\r^4$ (using equation (\ref{7})) and hence,
\be
&{}&\la v_{1\a}v_{2\b}\ra=\f{1}{2}\la v^2\ra\delta_{\a\b}-\f{5}{2}a\r^4\delta_{\a\b}+2a\r^2\r_\a\r_\b\nonumber\\
\Rightarrow&{}&\la (\pa_{1\tau}\pa_{1\gamma}v_{1\a})(\pa_{2\theta}\pa_{2\gamma}v_{2\b})\ra=-72a\delta_{\theta\tau}\delta_{\a\b}+24a\delta_{\b\theta}\delta_{\a\tau}\nonumber\\
&{}&\phantom{xxxxxxxxxxxxxxxxxxxxxx}+24a\delta_{\a\theta}\delta_{\b\tau}\nonumber\\
\Rightarrow&{}&\varepsilon_{\tau\a}\varepsilon_{\theta\b}\la (\pa_\tau\pa_\gamma v_{\a})(\pa_\theta\pa_\gamma v_{\b})\ra=-192a\\
\label{8-9}
\Rightarrow&{}&B_{\r\r}=-\f{\eta\r^4}{192\nu}
\label{9}
\en
In the equation (\ref{8-9}), we have put $\vec{\r}_1\approx\vec{\r}_2$, for these relations are assumed to be valid for arbitrarily small $\r$.
While writing the relation (\ref{9}), equation (\ref{0}) has been recalled.
This $B_{\r\r}$ is the second order correlation function for enstrophy cascade in dissipation range.
\section{THIRD ORDER CORRELATION FUNCTION FOR ENSTROPHY CASCADE IN INERTIAL RANGE}
\noindent Let's again define:
\ben
b_{\a\b,\gamma}\equiv\la v_{1\a}v_{1\b}v_{2\gamma}\ra
\enn
Invoking homogeneity and isotropy once again along with the symmetry in the first pair of indices, one may write the most general form of the third rank Cartesian tensor for this case as
\be
b_{\a\b,\gamma}&=&C(\r)\delta_{\a\b}\r^o_\gamma+D(\r)(\delta_{\gamma\b}\r^o_\a+\delta_{\a\gamma}\r^o_\b)\nonumber\\
&{}&+F(\r)
\r^o_\a\r^o_\b\r^o_\gamma
\label{10}
\en
where, $C$, $D$ and $F$ are functions of $\r$.
Yet again, incompressibility dictates:
\be
\f{\pa}{\pa_{2\gamma}}b_{\a\b,\gamma}=\f{\pa}{\pa_{\gamma}}b_{\a\b,\gamma}=0\nonumber\\
\Rightarrow C'\delta_{\a\b}+\f{C}{\r}\delta_{\a\b}+\f{2D}{\r}\delta_{\a\b}+\f{2D'}{\r^2}\r_\a\r_\b-\f{2D}{\r^3}\r_\a\r_\b\nonumber\\+\f{F'}{\r^2}\r_\a\r_\b+\f{F}{\r^3}\r_\a\r_\b=0
\label{11}
\en
Putting $\a=\b$ in equation (\ref{11}) one gets:
\be
2C+2D+F=\f{\textrm{constant}}{\r}=0
\label{11-12}
\en
where, it as been imposed that $b_{\a\b,\gamma}$ should remain finite for $\r=0$.
Again, using equation (\ref{11}), putting $\a\ne\b$ and manipulating a bit one gets:
\be
D=-\f{1}{2}(\r C'+C)
\label{12}
\en
using which in relation (\ref{11-12}), one arrives at the following expression for $F$:
\be
F=\r C'-C
\label{13}
\en
Defining
\be
B_{\a\b\gamma}&\equiv&\la(v_{2\a}-v_{1\a})(v_{2\b}-v_{1\b})(v_{2\gamma}-v_{1\gamma})\ra\nonumber\\
&=&2(b_{\a\b,\gamma}+b_{\gamma\b,\a}+b_{\a\gamma,\b})
\label{13-14}
\en
and putting relations (\ref{12}) and (\ref{13}) in the equation (\ref{13-14}) and using relation (\ref{10}), one gets:
\be
&{}&B_{\a\b\gamma}=-2\r C'(\delta_{\a\b}\r^o_\gamma+\delta_{\gamma\b}\r^o_\a+\delta_{\a\gamma}\r^o_\b)\nonumber\\
&{}&\phantom{xxxxxxx}+6(\r C'-C)\r^o_\a\r^o_\b\r^o_\gamma\\
\label{14}
\Rightarrow&{}&S_3\equiv B_{\r\r\r}=-6C
\label{15}
\en
which along with relations (\ref{12}), (\ref{13}) and (\ref{10}) yields the following expression:
\be
&{}&b_{\a\b,\gamma}=-\f{S_3}{6}\delta_{\a\b}\r^o_\gamma+\f{1}{12}(\r S'_3+S_3)(\delta_{\gamma\b}\r^o_\a+\delta_{\a\gamma}\r^o_\b)\nonumber\\
&{}&\phantom{xxxxxxxx}-\f{1}{6}(\r S'_3-S_3)\r^o_\a\r^o_\b\r^o_\gamma
\label{16}
\en
Navier-Stokes equation suggests:
\be
\f{\pa}{\pa t}v_{1\a}=-v_{1\gamma}\pa_{1\gamma}v_{1\a}-\pa_{1\a}p_1+\nu\pa_{1\gamma}\pa_{1\gamma}v_{1\a}
\label{17}\\
\f{\pa}{\pa t}v_{2\b}=-v_{2\gamma}\pa_{2\gamma}v_{2\b}-\pa_{2\b}p_2+\nu\pa_{2\gamma}\pa_{2\gamma}v_{2\b}
\label{18}
\en
multiplying equations (\ref{17}) and (\ref{18}) with $v_{2\b}$ and $v_{1\a}$ respectively and adding subsequently, one gets the following:
\be
\f{\pa}{\pa t}\la v_{1\a}v_{2\b}\ra&=&-\pa_{1\gamma}\la v_{1\gamma}v_{1\a}v_{2\b}\ra-\pa_{2\gamma}\la v_{2\gamma}v_{1\a}v_{2\b}\ra\nonumber\\
&{}&-\pa_{1\a}\la p_1v_{2\b}\ra-\pa_{2\b}\la p_2v_{1\a}\ra+\nu \pa_{1\gamma}\pa_{1\gamma}\la v_{1\a}v_{2\b}\ra\nonumber\\
&{}&+\nu \pa_{2\gamma}\pa_{2\gamma}\la v_{1\a}v_{2\b}\ra
\label{19}
\en
Due to isotropy, the correlation function for the pressure and velocity ($\la p_1\vec{v}_2\ra$) should have the form $f(\r)\vec{\r}/|\vec{\r}|$.
But since, $\pa_{\a}\la p_1{v}_{2\a}\ra=0$ due to solenoidal velocity field $f(\r)\vec{\r}/|\vec{\r}|$ must have the form $\textrm{constant}\times(\vec{\r}/|\vec{\r}|^3)$ that in turn must vanish to keep correlation functions finite even at $\r=0$.
Thus, equation (\ref{19}) can be written as:
\be
\f{\pa}{\pa t}b_{\a\b}=\pa_{\gamma}(b_{\a\gamma,\b}+b_{\b\gamma,\a})+2\nu \pa_\gamma\pa_\gamma b_{\a\b}
\label{20}
\en
Using equations (\ref{6}) and (\ref{16}), one can rewrite equation (\ref{20}) as:
\be
&{}&\f{1}{2}\f{\pa}{\pa t}\la v^2\ra-\f{1}{2}\f{\pa}{\pa t}B_{\r\r}=\nu \pa_\gamma\pa_\gamma\la v^2\ra+\f{1}{6\r^3}\f{\pa}{\pa \r}\left(\r^3B_{\r\r\r}\right)\nonumber\\
&{}&\phantom{xxxxxxxxxxxxxxxxx}-\f{\nu}{\r}\f{\pa}{\pa \r}\left(\r\f{\pa B_{\r\r}}{\pa \r}\right)
\label{21}
\en
As we are interested in the enstrophy cascade, the first term in the R.H.S. is zero due to homogeneity and the first term in the L.H.S. is zero because of energy remains conserved in 2-D turbulence in the inviscid limit (and it is the high Reynolds number regime that we are interested in); it cannot be dissipated at smaller scales.
Also, as we are interested in the forward cascade which is dominated by enstrophy cascade, on the dimensional grounds in the inertial region $B_{\r\r}$ (as it may depend only on $\eta$ and $\r$) may be written as:
\be
\f{\pa}{\pa t}B_{\r\r}=A\eta\r^2
\label{22}
\en
where $A$ is a numerical proportionality constant.
Hence, using the relation (\ref{22}), the equation (\ref{21}) reduces to the following differential equation:
\be
\f{1}{6\r^3}\f{\pa}{\pa \r}\left(\r^3B_{\r\r\r}\right)=\f{\nu}{\r}\f{\pa}{\pa \r}\left(\r\f{\pa B_{\r\r}}{\pa \r}\right)-\f{A}{2}\eta\r^2
\label{23}
\en
which when solved using relation(\ref{15}) in the limit of infinite Reynolds number ($\nu\rightarrow 0$), one gets
\be
B_{\r\r\r}=-\f{A\eta}{2}\r^3
\label{24}
\en
The relation (\ref{24}) is the expression for the third order correlation function of the isotropic and homogeneous 2-D decaying turbulence in the inertial range of the forward cascade.
The fact that we could not determine an exact value for $A$ is what is disturbing.
Bernard's procedure leads to `$A$'$=-1/4$.
One knows that in two-dimensional turbulence the energy spectrum in enstrophy cascade regime is of the form $K{\eta}^{2/3}k^{-3}\{\ln(k/k_{1})\}^{-1/3}$ (where $K$ is a numerical constant and $k\gg k_1$) and hence, recalling that the integral of energy spectrum over $k$ is directly related to $B_{\alpha\beta}$, in the relation (\ref{22}), the dimensional form of $\dot{B}_{\r\r}$ should have a logarithmic correction.
This would immediately lead to a logarithmic correction in $B_{\r\r\r}$, making the existence of `$A$' as a constant meaningless.
It is possible that the difference between the Bernard's and our approaches shows a basic difference between a forced turbulence and a decaying turbulence which comes to the fore in the example being dealt with.
\section{THIRD ORDER CORRELATION FUNCTION FOR ENERGY CASCADE IN INERTIAL RANGE}
Suppose in the homogeneous isotropic fully-developed turbulence in 2-D, energy is being supplied and the mean rate of injection of energy per unit mass is denoted by $\varepsilon$.
Let us concentrate on the inverse energy cascade.
Then technically we have to proceed as in the previous section and on doing so one would arrive at the differential equation (\ref{21}); only that now the arguments would differ.
In the larger scales viscosity is not as significant and anyway we shall be interested in the infinite Reynolds number case which would mean that the last term in the R.H.S. of equation (\ref{21}) would go to zero.
One obviously would also set $\f{1}{2}\f{\pa}{\pa t}\la v^2\ra=\varepsilon$ and lets assume $\f{\pa}{\pa t}B_{\r\r}\approx 0$ in the inverse cascade regime, justification of which can be sought from the fact that the ultimate result that is obtained has been experimentally and numerically verified.
So we are left with the following differential equation:
\be
&{}&\f{1}{6\r^3}\f{\pa}{\pa \r}\left(\r^3B_{\r\r\r}\right)=\varepsilon\\
\R&{}&B_{\r\r\r}=+\f{3}{2}\varepsilon\r
\label{energy_cascade_S3}
\en
where in the last step the integration constant has been set to zero to prevent $B_{\r\r\r}$ from blowing up at $\r=0$.
\section{SECOND ORDER VORTICITY CORRELATION FUNCTION IN DISSIPATION RANGE OF ENSTROPHY CASCADE}
Let us define:
\be
&{}&W\equiv\la\omega_1\omega_2\ra
\label{25}\\
\textrm{and,}\phantom{xxx}&{}&\Omega\equiv\la(\omega_2-\omega_1)(\omega_2-\omega_1)\ra
\label{26}
\en
Due to homogeneity, $\Omega$ may be expressed as:
\be
\Omega=2\la\omega^2\ra-2W
\label{27}
\en
In the dissipation range: $v\sim\r^2$ so, $\omega\sim\r$ and hence we may, choosing a proportionality constant $b$ (say), assume:
\be
\Omega=b\r^2
\label{28}
\en
Using relations (\ref{25}),(\ref{27}) and (\ref{28}), one gets:
\be
&{}&\la\omega_1\omega_2\ra=\la\omega^2\ra-\f{b}{2}\r^2\\
\R&{}&\la(\pa_{1\a}\omega)(\pa_{2\a}\omega)\ra=2b
\label{29}
\en
But we know,
\be
\eta=\nu\la(\pa_{\a}\omega)(\pa_{\a}\omega)\ra
\en
So, relation (\ref{29}) would yield:
\be
\eta=2\nu b
\label{30}
\en
where, we have put $\vec{\r}_1\approx\vec{\r}_2$ in the relation (\ref{29}), for, being in the dissipation range, these relations are assumed to be valid for arbitrarily small $\r$.
Using relations (\ref{28}) and (\ref{30}), one arrives at a experimentally verifiable result for two-point second order vorticity correlation function:
\be
\Omega=\f{\eta}{2\nu}\r^2
\label{31}
\en
\section{THIRD ORDER MIXED CORRELATION FUNCTION IN INERTIAL RANGE OF ENSTROPHY CASCADE}
We start by defining a two-point third order mixed correlation tensor in inertial range:
\be
&{}&\Omega_\b\equiv\la(v_{2\b}-v_{1\b})(\omega_2-\omega_1)(\omega_2-\omega_1)\ra
\label{32}\\
\R&{}&\Omega_\b=2M_{\b}+4W_{\b}
\label{33}
\en
where,
\be
&{}&W_{\b}\equiv\la v_{1\b}\omega_1\omega_2\ra
\label{34}\\
\textrm{and,}\phantom{xxx}&{}&M_{\b}\equiv\la\omega_1\omega_1v_{2\b}\ra
\label{35}
\en
Let due to isotropy and homogeneity, we can write following form for $M_{\b}$:
\be
&{}&M_{\b}=M(\r)\r_{\b}^o
\label{36}\\
\R&{}&\f{\pa}{\r_{2\b}}M_{\b}=\la\omega_1\omega_1\pa_{2\b}v_{2\b}\ra=0
\label{37}\\
\R&{}&\f{\pa}{\pa\r}M({\r})+\f{M({\r})}{\r}=0\\
\R&{}& M({\r})=\f{\textrm{constant}}{\r}=0
\label{38}
\en
In the relation (\ref{37}), we are assuming incompressibility and in writing the relation (\ref{38}) we have taken into account the fact that $M_{\b}$ should remain finite when $\r=0$.
Relations (\ref{36}) and (\ref{38}) imply that:
\be
M_{\b}=0
\label{39}
\en
using which in the relation (\ref{33}), we get:
\be
\Omega_{\b}=4W_{\b}
\label{40}
\en
From the equations (\ref{17}) and (\ref{18}), we may write respectively:
\be
\f{\pa}{\pa t}\omega_{1}=-v_{1\gamma}\pa_{1\gamma}\omega_{1}+\nu\pa_{1\gamma}\pa_{1\gamma}\omega_{1}
\label{41}\\
\f{\pa}{\pa t}\omega_{2}=-v_{2\gamma}\pa_{2\gamma}\omega_{2}+\nu\pa_{2\gamma}\pa_{2\gamma}\omega_{2}
\label{42}
\en
Multiplying equations (\ref{41}) and (\ref{42}) by $\omega_2$ and $\omega_1$ respectively and adding subsequently, we get the following differential equation:
\be
\f{\pa}{\pa t}W=2\pa_{\b}W_{\b}+2\nu\pa_{\b}\pa_{\b}{W}
\label{43}
\en
where we have used the fact $\pa_{\b}=-\pa_{1\b}=\pa_{2\b}$.
Using relations (\ref{27}) and (\ref{40}) in the equation ({\ref{43}}), one gets for the inertial range for the enstrophy cascade in homogeneous, isotropic and fully-developed freely decaying turbulence in 2-D in the infinite Reynolds number limit ({\it{i.e.}}, $\nu \rightarrow 0$) following differential equation:
\be
&{}&\f{\pa}{\pa t}\la\omega^2\ra-\f{1}{2}\f{\pa}{\pa t}\Omega=\f{1}{2\r}\f{\pa}{\pa\r}\left(\r\Omega_{\r}\right)
\label{44}\\
\R&{}&\Omega_{\r}=-2\eta\r
\label{45}
\en
In getting relation ({\ref{45}}) from the equation (\ref{44}), we have used the facts: $\f{1}{2}\f{\pa}{\pa t}\la\omega^2\ra=-\eta$ and $\f{1}{2}\f{\pa}{\pa t}\Omega\approx 0$ as it may be supposed that the value of $\Omega$ varies considerably with time only over an interval corresponding to the fundamental scale of turbulence and in relation to local turbulence the unperturbed flow may be regarded as steady which mean that for local turbulence one can afford to neglect $\f{\pa}{\pa t}\Omega$ in comparison with the enstrophy dissipation rate $\eta$.
 This result (relation (\ref{45})) has gained importance by serving as the starting point in deriving various rigorous inequalities for short-distance scaling exponents in 2-D incompressible turbulence\cite{Eyink}.
\section{CONCLUSION}
The Kolmogorov-Landau approach has been invoked in 2-D homogeneous isotropic decaying fluid turbulence to arrive at the various correlation functions earlier obtained using different methods.
Also, some experimentally verifiable correlation functions in the dissipation range have been derived, which to the best of our knowledge, have not been reported anywhere in the literature yet.
The results derived here are `exact' (though not rigorous) something which is a far cry in the literature on turbulence.
Besides, we have shown that the approach we have adopted here possibly highlights the fact that the logarithmic corrections in 2-D turbulence may hinder the proof of ``one-eighth law'' in the enstrophy cascade region, and thereby showcases the handicap of the Kolmogorov-Landau approach in 2-D fluid turbulence.\\ \\
The author would like to acknowledge his supervisor Prof. J. K. Bhattacharjee for the helpful discussions. Also, CSIR (India) is gratefully acknowledged for awarding fellowship to the author. Ayan's help in providing the author with relevant scholarly articles is heartily appreciated.

\end{document}